\documentclass[12pt]{iopart}

\usepackage{hyperref}
\usepackage[dvips]{graphicx}

\def\v#1{{\bf#1}}
\def\be{\begin{equation}}
\def\ee{\end{equation}}
\def\bea{\begin{eqnarray}}
\def\eea{\end{eqnarray}}

\newcommand{\bfkappa}{\mbox{\boldmath$\kappa$\unboldmath}}
\def\ie{{\it i.e.\,}}
\def\etal{{\it et al.   }}

\def\<{\langle}
\def\>{\rangle}

\begin{document}

\title{Playing relativistic billiards beyond graphene}
\author{E. Sadurn\'i $^{1,2}$, T. H. Seligman $^{2,3}$ and F. Mortessagne $^4$}

\address{$^1$Institut f\"ur Quantenphysik, Ulm Universit\"at, Albert-Einstein Allee 11 89081 Ulm - Germany.}

\address{$^2$Centro Internacional de Ciencias, Universidad Nacional Aut\'onoma de M\'exico, Av. Universidad s/n, 62210 Morelos, M\'exico.}

\address{$^3$Instituto de Ciencias F\'{\i}sicas,
Universidad Nacional Aut\'onoma de M\'exico, Av. Universidad s/n, 62210 Morelos, M\'exico.}

\address{$^4$Laboratoire de Physique de la Mati{\`e}re Condens{\'e}e,
Universit{\'e} de Nice-Sophia Antipolis, CNRS, UMR 6622
Parc Valrose - 06108 Nice cedex 2 - France.}

\eads{ \mailto{esadurni@uni-ulm.de},
\mailto{seligman@fis.unam.mx},
\mailto{fabrice.mortessagne@unice.fr}}

\begin{abstract}

The possibility of using hexagonal structures in general and graphene in particular to
emulate the Dirac equation is the basis of our considerations. We show that Dirac oscillators
with or without restmass can be emulated by
distorting a tight binding model on a hexagonal structure. In a quest to make a toy model for
such relativistic equations we first show that a hexagonal lattice of attractive potential wells
would be a good candidate. First we consider the corresponding one-dimensional model giving rise to a one-dimensional Dirac oscillator, and then construct explicitly the deformations needed in the two-dimensional case.
Finally we discuss, how such a model can be implemented as an electromagnetic billiard using arrays of dielectric resonators
between two conducting plates that ensure evanescent modes outside the resonators for transversal electric modes, and describe an appropriate experimental setup.

\end{abstract}

\submitto{\NJP}

\maketitle

\section{Introduction}

Dirac operators play a central role in relativistic quantum dynamics, 
from the early work of Dirac and the exact solutions for the hydrogen 
atom to later work including quantum chromodynamics and the Dirac oscillator \cite{dirac, verbaarschot1, mosh}.
Based on early work of Wallace \cite{wallace}, there has been much recent work centered around the
fact that the mean field theory of grapene is well represented by 
the free Dirac equation near the edges of the first Brillouin zone, \ie at the center of the band with so-called Dirac points around which linear dispersion relations hold \cite{novoselov, semenoff, Cas09}. Simulations of this kind of situation are ongoing at various labs using hexagonal and occasionally
triangular arrays in single particle problems or classical waves, including acoustics \cite{acoustic}, microwaves and photonic crystals \cite{Lou05} as well as true quantum simulations in nanostructures \cite{sahin, sevicli, ouyang, thomas}. We wish to recall, that any lattice with coordination number three will yield points with approximate linear dispersion relations similar to Dirac points, but we need a tight bindig situation to guarantee isotropic cones and a hexagonal structure to introduce an additional discrete degree of freedom for the small and large component of the effective spinor. This becomes very transparent as we note, that using a hexagonal structure made up of two different triangular structures, we obtain a finite gap in the spectrum, i.e a Dirac equation for a massive particle, a situation that would correspond to a B-N (Boron Nitride) lattice.

In this paper we will adress the task, to find similar analoga for Dirac operators, that describe situations other than the free particle. Formally we shall assume arrays of potential wells, that can hold exactly one bound state. In between the wells these states decay exponentially. As the overlap of functions of two wells describes the coupling, this implies to a good approximation a tight binding system. In this framework in principle one and two - dimensional Dirac type problems can be formulated, and in some cases exactly solved. We shall here focus on the Dirac oscillator as introduced by Marcos Moshinsky; this system is frequently used \cite{nouicer, decastro, alhaidari, mielnik, castanos, nosotros, yoconmarcos}, and is particularly simple to implement. Other solvable problems, such as gyroscopes \cite{yo} and the Coulomb problem as well as systems with random potentials will be touched upon in the conclusions. 

In particular, the one and two dimensional Dirac oscillators have been considered recently in the context of both relativistic quantum mechanics and quantum optics \cite{bermudez, bermudez2, nosotros2}. Their importance as paradigmatic integrable models is well established and their realization in the context of tight binding models is desirable. We want to stress that the dimensionality in our examples plays a crucial role in their algebraic properties and spectra. As we shall see, the exceptional infinite degeneracy of the two dimensional Dirac oscillator (as opposed to the finite degeneracy in one dimension) is intimately related to its specific realization on the lattices described above. It is therefore interesting to consider both dimensionalities, as they exhibit clear differences. It remains as an open question, whether three dimensional relativistic wave equations can be emulated on a three dimensional lattice. 

The description of such systems will depend on a massive distortion of the hexagonal lattice.
This causes probably prohibitive obstacles to any implementation in terms of mean fields of graphene related structures. For recent theoretical work dealing with deformations on carbon sheets see \cite{lewenkopf}. For classical wave models on the other hand such distortions can be implemented 
and we shall discuss specifically how this is done in what we usually call a microwave billiard.
In two dimensions such systems yield scalar wave equations, and can readily be interpreted as single particle quantum systems \cite{stoeckmann}.

Microwave billiards have been used to simulate a wide variety of phenomena including quantum chaos
\cite{doron, Gra92, So95, Sto02}, scattering from open billiards \cite{dembowski, darmstadt}, transport phenomena \cite{transport, transport2}, fidelity decay \cite{schaefer1, schaefer2} and disordered systems \cite{Lau07, correlated}, as well as recent work on Dirac points \cite{Bar09}, \cite{darmstadt2}. This wide range of success combined with recent experiments with arrays of small dielectric micro wave resonators gives us good reason to hope for interesting results in orderd structures such as the ones we need. The evanescence of the cavity modes, that we intend to use, guarantees exponential decay and thus tight binding situations can be emulated.

In the next section we study a one dimensional crystal; this will serve us to fix notation and basic ideas
in a simpler context, that is not without relevance in itself. This toy model reveals Dirac-like hamiltonians in both periodic and deformed structures. In section 3 we analyze and reproduce similar results for the two dimensional case with and without deformations. We also consider corrections to nearest neighbour interactions and obtain the form of energy surfaces beyond tight binding. In section 4 we make contact with experimental applications of our treatment by considering arrays of resonators in microwave cavities. The applicability of the tight binding model in this context is discussed. Finally, we draw some conclusions and give an outlook on a selection of other Dirac systems, that can be emulated. Useful results are included in the appendix.

\section{One-dimensional crystal}

In this section we start out with two lattices. First, we consider the Schroedinger equation with square wells supporting a single bound state, located in a periodic one dimensional array. Here, traditional aspects of the existing theory (\ie Bloch waves) are reviewed under an algebraic approach. The specific form of the localized wave functions is ignored, as it results to be irrelevant. An analogy between degeneracy points of the spectrum and a one dimensional Dirac point is revealed in this toy model. Near these points we shall cover a one-dimensional Dirac equation.

Once this is achieved, we proceed to deform the lattice from its periodic configuration by forcing a specific operator algebra, namely the one corresponding to harmonic oscillator ladder operators. As an outcome, a one dimensional Dirac oscillator shall be realized. 

\subsection{The Dirac point in one dimension}

\begin{figure}
\begin{center}
\includegraphics[scale=0.65]{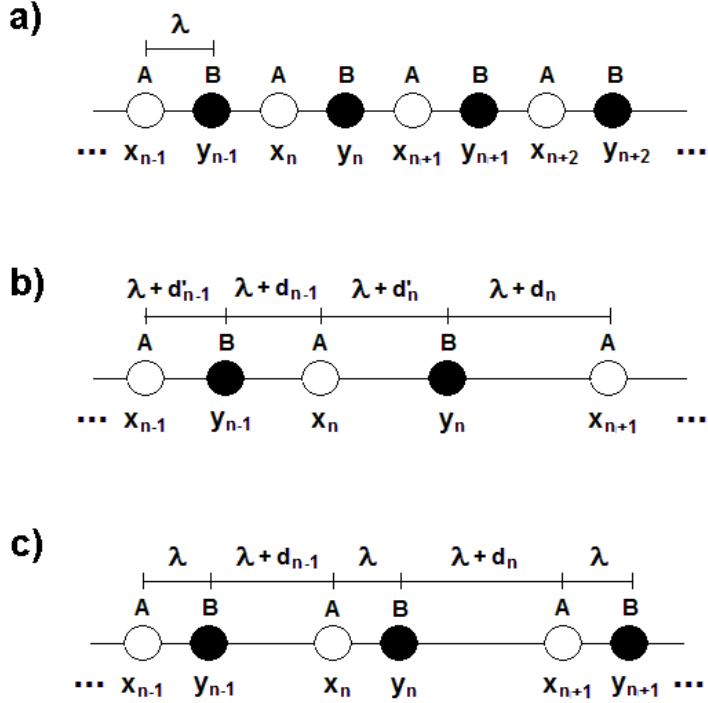}
\end{center}
\caption{\small{Configuration of potential wells (or resonators) on a one-dimensional lattice. Chain a) shows the periodic case. Chain b) corresponds to a general deformation of both sublattices. Chain c) shows the resulting deformation after imposing the harmonic oscillator algebra.}}
\label{fig1}
\end{figure}

We start by fixing some notation. In the following, we adopt natural units $\hbar=c=1$. A lattice consisting of two periodic sublattices is considered. Let $\lambda$ be the distance between neighbouring potential wells. They have the same period and are denoted as type A and type B. Each sublattice point can be labeled by an integer $n$ according to its position on the line, \ie $x_n$ for type A and $y_n$ for type B (see figure \ref{fig1}). The energy of the single level to be considered in the well is denoted by $\alpha$ and $\beta$ for type A and B respectively. The state corresponding to a particle at site $n$ of lattice A is denoted by $|n\>_A$ and the corresponding localized wave function is given by $\xi_A(x-x_n) = \<x|n\>_A$. For lattice B we define the wavefunction $\xi_B(x-y_n) = \<x|n\>_B$. Our hamiltonian is well approximated by a tight binding model if the overlap between localized wave functions can be neglected and the nearest neighbour coupling matrix element $\Delta$ is taken as the first off-diagonal element of the hamiltonian in the localized basis. 

For convenience we have split the lattice into two sublattices and we write the hamiltonian in the form 

\bea
H= \left( \begin{array}{cc} H_{AA}& H_{AB} \\ H_{BA}& H_{BB}\end{array} \right)
\label{1}
\eea
rather than in the usual tridiagonal form. The entries of each block are given by $H_{ij}^{n,m}= \<n| H_{ij} |m\>$ with $n,m$ integers. Thus, each block extends from $\<-\infty| H_{ij} |-\infty\>$ to $\<\infty| H_{ij} |\infty\>$. In this basis we write the hamiltonian as

\bea
H= \left( \begin{array}{cc} \begin{array}{cccc} \ddots & & & \\ & \alpha & & \\ &&\alpha& \\ &&& \ddots \end{array} & \begin{array}{cccc} \ddots & & & \\ & \Delta & \Delta & \\ &&\Delta&\Delta \\ &&& \ddots \end{array} \\ \begin{array}{cccc} \ddots &  & & \\ & \Delta &  & \\ &\Delta &\Delta& \\ &&\Delta& \ddots \end{array} & \begin{array}{cccc} \ddots & & & \\ & \beta & & \\ &&\beta& \\ &&& \ddots \end{array}\end{array} \right)
\label{2}
\eea
where the elements outside the indicated diagonals are all zero. Expression (\ref{2}) can be cast in terms of Pauli matrices $\sigma_3,\sigma_+ =\sigma_1+i\sigma_2, \sigma_-=\sigma_+^{\dagger}$ by defining $\Pi=H_{AB} \otimes \v 1$ and setting $M=(\alpha-\beta)/2, E_0=(\alpha+\beta)/2$. We have 

\bea
H= E_0 + \sigma_3 M + \sigma_+ \Pi + \sigma_- \Pi^{\dagger}
\label{4}
\eea
displaying explicitly the Dirac-like structure of the hamiltonian. To ensure the analogy between the Dirac hamiltonian and (\ref{4}), we consider the spectrum of $H$. We note that $\Pi$ has the remarkable properties

\bea
[\Pi,\Pi^{\dagger}]=0, \qquad \Pi \Pi^{\dagger} = \Delta (\Pi + \Pi^{\dagger}).
\label{5}
\eea
From these relations we may compute the spectrum by squaring $H$

\bea
(H-E_0)^2 = M^2 + \Pi \Pi^{\dagger}.
\label{6}
\eea
Note that Bloch's theorem manifests itself in the spectrum and eigenfunctions of $\Pi\Pi^{\dagger}$ as

\bea
\Pi \phi_k = \Delta (1+e^{i2\pi \lambda k}) \phi_k, \quad \Pi \Pi^{\dagger} \phi_k = \Delta^2 |1+e^{i2\pi \lambda k}|^2 \phi_k
\label{7}
\eea
with 

\bea
\phi_k = \sum_{n=-\infty}^{\infty} e^{i2\pi \lambda k n} |n\> = \left(\begin{array}{c} \vdots\\ e^{i2\pi \lambda k n} \\ e^{i2\pi \lambda k (n+1)} \\ \vdots \end{array} \right).
\label{8}
\eea
Therefore energies and eigenfunctions of $H$ are given by

\bea
E(k) = E_0 \pm \sqrt{\Delta^2|1+e^{i2\pi \lambda k}|^2 + M^2}, \quad \psi^{\pm} = N \left(\begin{array}{c} \phi_k \\ \frac{\pm E(k) - E_0 - M}{\Delta(1+e^{i2\pi \lambda k})} \phi_k \end{array} \right),
\label{9}
\eea
where $N$ is a normalization constant. When $M=0$ \ie when the sites A and B are equal, we can find conical points of the form $k_0= \pm 1/(2\lambda)$. Expanding around these points we find the usual expressions for the relativistic energies of Dirac particles with momentum $\kappa=k-k_0$, effective speed of light $\Delta$:

\bea
E(\kappa)= E_0 \pm \sqrt{\Delta^2 \kappa^2 + M^2}.
\label{10}
\eea
This is valid also for non-zero rest energy $M$ in the case of different lattices. Then we find a gap in the spectrum. The eigenfunctions satisfy the Dirac equation in momentum space

\bea
(E-E_0) \psi^{\pm}(\kappa) = [ \sigma_1 \kappa + \sigma_3 M ]\psi^{\pm}(\kappa).
\label{11}
\eea
In the next subsection we shall proceed to show that we can obtain the Dirac oscillator by deforming the double lattice.

\subsection{One-dimensional Dirac oscillator and lattice deformations}

Using the notation of the previous section, we modify the energy operator by deforming the lattice of potential wells. This implies abandoning the periodic structure and leads to a site dependence of the couplings in the corresponding tight binding model. We denote by $\Delta_{n,n+1}$ the coupling between sites at $y_n$ and $x_{n+1}$, while $\Delta_{n,n}$ denotes the coupling between sites $x_n$ and $y_n$. These couplings are approximately proportional to the overlap between neighbouring sites. They decay exponentially as a function of the separation distance between the potential wells, \ie

\bea
\Delta_{n,n+1} = \Delta e^{-d_n / \Lambda}, \qquad \Delta_{n,n} = \Delta e^{-d'_n / \Lambda}
\label{12}
\eea
where $d_n$ and $d'_n$ are the deviations from the periodic configuration, \ie $d_n + \lambda = |y_{n+1}-x_{n}|$, $d'_n + \lambda = |y_{n}-x_{n}|$. When $d_n=d'_n=0$, the periodic configuration is recovered (see figure). The length $\Lambda$ is the penetration depth into the classically forbidden region for the wave function of the single well. We expect a modification of the operators $\Pi,\Pi^{\dagger}$ caused by the change in the position of the wells. One should keep in mind that such deformations have the effect of breaking the periodic symmetry of the system and Bloch wave functions cease to be useful. However, if one finds an exactly solvable model for a deformation which is continuously connected to the periodic configuration, then the corresponding solutions could be considered to constitute a generalization of Bloch's waves.

The algebraic properties of observables in hamiltonians have a clear connection with integrability and exact solvability. Therefore, the simplest way to extend our hamiltonian is by replacing the operators $\Pi, \Pi^{\dagger}$ by $a,a^{\dagger}$, such that their algebraic properties are those of known solvable systems. In particular, we propose the harmonic oscillator algebra, since it is a paradigmatic example \cite{moshbook}. The hamiltonian (\ref{4}) becomes 

\bea
H= E_0 + \sigma_3 M + \sigma_+ a + \sigma_- a^{\dagger}.
\label{12.1}
\eea
We require that such extensions reduce to the periodic case in some {\it free\ }limit. We propose then

\bea
a = \left( \begin{array}{cccc} \ddots & & & \\ & \Delta_{n,n} & \Delta_{n,n+1} & \\ &&\Delta_{n+1,n+1}&\Delta_{n+1,n+2} \\ &&& \ddots \end{array} \right)
\label{13}
\eea
and impose the condition

\bea
[a,a^{\dagger}] = \omega \Delta = {\rm constant}
\label{14}
\eea
where $\omega$ stands for a frequency. If $\omega \Delta = 0$ we recover the algebra of $\Pi,\Pi^{\dagger}$ (Bloch limit). Computing the commutator in (\ref{14}), one finds the conditions

\bea
\Delta_{n,n} = \Delta,  \qquad \Delta^2_{n+1,n+2}- \Delta^2_{n,n+1}= \omega \Delta
\label{15}
\eea
where the second equality is a recurrence equation. The corresponding solution is 

\bea
\Delta_{n,n+1}= \sqrt{\Delta^2-n\omega \Delta}, \qquad n \in \v Z
\label{16}
\eea 
and thus

\bea
a = \left( \begin{array}{cccc} \ddots & & & \\ & \Delta & \sqrt{\Delta^2-n\omega \Delta} & \\ &&\Delta&\sqrt{\Delta^2-(n-1)\omega \Delta} \\ &&& \ddots \end{array} \right)
\label{17}
\eea
Imposing the conditions (\ref{12}) in this solution we find

\bea
\Delta e^{-d'_n / \Lambda} = \Delta \\ \nonumber
\Delta e^{-d_n / \Lambda} = \sqrt{\Delta^2-n\omega \Delta } \in \v R \\ \nonumber 
\label{18}
\eea
The first condition implies $d'_n=0$ (a chain of dimers). The second condition relates the displacement between resonators $d_n$ with the corresponding couplings.

For convenience we impose a third condition, namely the inequality

\bea
d_n > d_{n-1}
\label{18.1}
\eea
 This condition ensures that the displacement $d_n$ increases monotonically as a function of the distance from an arbitrarily chosen origin. Combining it with the second condition this implies a finite grid, which in turn leads to a finite Hilbert space of dimension $n_{max}+1$ with $n_{max}= [|\frac{\Delta}{\omega}|]$. The operator $a$ takes the finite matrix form

\bea
a'=\left( \begin{array}{ccc}   \Delta & \sqrt{\Delta^2-n_{max}\omega \Delta} & \\ &\Delta&\sqrt{\Delta^2-(n_{max}-1)\omega \Delta} \\ && \ddots \end{array} \right).
\label{19}
\eea
In principle other non-monotonical choices of scaling $d_n$ are possible and lead to other finite or infinite arrays. For our modeling purposes finite arrays are advantageous, and in any case our hamiltonian will emulate a Dirac oscillator only in the vicinity of zero energy (energies near $E_0$) as we will see later.

We now have to check the commutation relations between $a'$ and $a'^{\dagger}$

\bea
[a',a'^{\dagger}] &=& \omega \Delta \v 1 + [\Delta^2-(n_{max}+1)\omega \Delta] \left( \begin{array}{cccc}   1 & & &\\ &0& &\\ &&0& \\ &&&\ddots \end{array} \right) \\ \nonumber
&=& \omega \Delta \v 1 + O(1/n_{max}).
\label{20}
\eea
Note that the correction term is of order $1/n_{max}$ as the principal term is accompanied by the identity, while the correction acts only in the first component of this basis. The prefactor is of order 1 since $ \Delta / \omega  \simeq [| \Delta / \omega |] + 1$. The distortion is

\bea
d_n = \Lambda \log{\left( \frac{\Delta^2}{\Delta^2-n \omega \Delta} \right)}, \qquad  0\leq n \leq n_{max}
\label{21}
\eea
Now we can replace $a'$ by $a$ and return to the hamiltonian (\ref{12.1}) for the finite system. Our construction allows to compute the spectrum and eigenfunctions up to a shift smaller than unity. As we shall see, they correspond to those of the Dirac oscillator \cite{mosh}. We consider eigenvectors $\phi_n$ of the number operator such that $a^{\dagger}a \, \phi_n = (\omega \Delta) n \phi_n$, $ a^{\dagger} \phi_n = \sqrt{\omega \Delta (n+1)}\phi_{n+1} $ and $a \phi_0 = 0$. The hamiltonian has an integral of the motion given by the operator $I= a^{\dagger} a + \frac{1}{2} \omega \Delta (\sigma_3 + 1)$ with eigenstates $ \phi_n |+\>, \phi_{n+1} |-\> $. In this basis, $H-E_0$ is reduced to $2 \times 2$ blocks of the form

\bea
\left( \begin{array}{cc} M &  \sqrt{\omega \Delta(n+1)} \\ \sqrt{\omega \Delta(n+1)} & -M \end{array} \right).
\label{21.1}
\eea
These blocks can be diagonalized, leading to the energies

\bea
E(n) = E_0 \pm \sqrt{\omega \Delta (n+1) + M^2}, \qquad  0 \leq n \leq n_{max}.  
\label{21.2}
\eea
An additional $1 \times 1$ block is due to the singlet $\psi_{0}=\phi_0 |-\>$, leading to $H\psi_0 =(E_0-M)\psi_0 $. The eigenfunctions corresponding to the doublets are obtained in the form   

\bea
\fl \psi_{n+1}^{\pm} = N \left( \phi_{n} | + \> + \frac{\pm (E(n) - E_0) - M}{\sqrt{\omega \Delta (n+1)}} \phi_{n+1} | - \> \right),\qquad  0 \leq n \leq n_{max}, \\ \nonumber
\label{24}
\eea
where $N$ is a normalization constant. The specific form of $\phi_n$ is given in the appendix. 

In the next section we shall see that a similar construction occurs very naturally if we deform hexagonal lattices in two dimensions, which in the undeformed case are closely related to the mean field theory of graphene and boron nitride sheets. 

\section{Two-dimensional crystal}

The concepts given in the last section are now extended to two dimensions. We shall consider hexagonal structures that emulate two-dimensional Dirac equations and can emulate the mean field of graphene or Boron-Nitride near the gap at the center of the usual band. We shall use the same algebraic strategy to derive spectra and a possible extension through deformations, namely the two-dimensional Dirac oscillator.

\subsection{Hexagonal lattice}

\begin{figure}
\begin{center}
\includegraphics[scale=0.35]{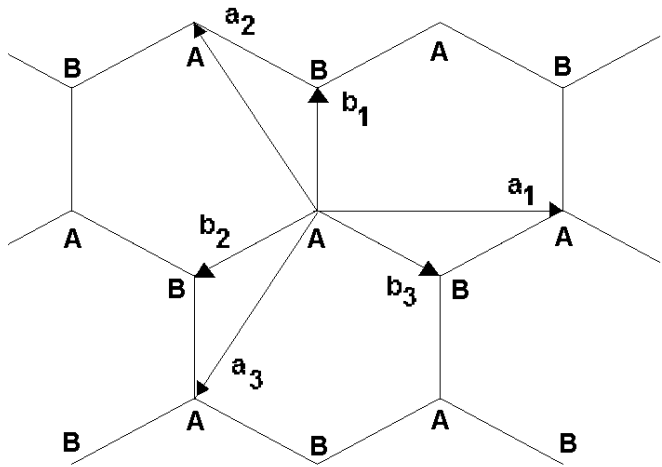}
\end{center}
\caption{\small{Vectors in a two dimensional hexagonal lattice. The vectors $\v b_i$ connect sublattice A with sublattice B. The vectors $\v a_i$ connect points in the same sublattice.}}
\label{hexa}
\end{figure}

Let us fix the notation for this system. The honeycomb lattice is divided in two triangular sublattices, one of them generated by the set of vectors $\v a_1=(3 \sqrt{2},0),\v a_2=(-3/\sqrt{2},3/2), \v a_3=(3/\sqrt{2},-3/2)$ (type A) while the other sublattice is obtained by adding the vectors $\v b_1=(0,1), \v b_2=(-3/\sqrt{2},-1/2), \v b_3=(3/\sqrt{2},-1/2)$. These vectors are given in arbitrary units (see figure \ref{hexa}). We denote the linear combinations of $\v a_i$ by $\v A$, where $\v A$ is a vector parametrizing the points of sublattice A. For sublattice B we use the vector parameter $\v A + \v b_1$. The position vectors $\v r_A, \v r_B$ of the periodic lattices are obtained by introducing the factor $\lambda$. For periodic arrays this means $\v r_A=\lambda \v A, \v r_B =\lambda \v B$. In further considerations this notation will be useful, since deformed lattices admit a parametrization by vectors $\v A, \v B$, but the corresponding position vectors $\v r_A, \v r_B$ become more complicated functions of $\v a_i, \v b_i$. The state vectors for individual potential wells on grid A shall be denoted by $| \v A \>$, giving wave functions of individual wells as $\xi_A(\v r - \v r_A)=\< \v r | \v A \>$. For grid B we use $| \v A + \v b_1\>$. As before, we consider different energies for the wells on grids A and B. 

Similarly as in the one-dimensional case, the hamiltonian in the tight binding approximation is constructed as

\bea
H &=& \alpha \sum_{\v A} |\v A \> \< \v A | + \beta \sum_{\v A} |\v A + \v b_1 \> \< \v A + \v b_1 |   \\ \nonumber
&+& \sum_{\v A, i=1,2,3} \Delta \left( |\v A \> \< \v A + \v b_i | + |\v A + \v b_i \> \< \v A | \right)
\label{28}
\eea
where the first two terms indicate the total on-site energy on grids A and B respectively, while the last sum indicates the nearest neighbour interaction with coupling strength $\Delta$. We shall analyze this system by considering again a subdivision of the Hilbert space according to sublattices A and B. Due to the coordination number in this lattice, the matrix representation in section 2 is no longer feasible. However, we may construct the usual Pauli operators through the definitions

\bea
\sigma_{+} = \sum_{\v A} |\v A \> \< \v A + \v b_1 |, \qquad \sigma_{-} = \sigma_{+}^{\dagger} \\ \nonumber
\sigma_3 = \sum_{\v A} |\v A \> \< \v A | - |\v A + \v b_1 \> \< \v A + \v b_1 |, \quad
\v 1 = \sum_{\v A} |\v A \> \< \v A | + |\v A + \v b_1 \> \< \v A + \v b_1 |,
\label{29}
\eea
while the operators $\Pi, \Pi^{\dagger}$ are defined through

\bea
\Pi = \sum_{\v A, i} \Delta \left( |\v A \> \< \v A + \v b_i - \v b_1 | + |\v A + \v b_1 \> \< \v A + \v b_i | \right).
\label{29.1}
\eea
They have the algebraic properties

\bea
[\sigma_3,\sigma_{\pm}] = \pm 2 \sigma_{\pm}, \quad [\Pi,\Pi^{\dagger}]=0, \quad [\Pi,\sigma_i] = 0
\label{30}
\eea
With $M$ and $E_0$ given as in section 2, we obtain the hamiltonian 

\bea
H= E_0 + M\sigma_3 + \sigma_+ \Pi + \sigma_- \Pi^{\dagger}
\label{33}
\eea
and

\bea
(H-E_0)^2 = M^2 + \Pi \Pi^{\dagger}
\label{33.1}
\eea
Using Bloch waves, we have eigenvectors of the form \cite{Blochhexagonal}

\bea
\phi^{a}_{k} = \sum_{\v A, i} e^{-i 2 \pi \lambda \v k \cdot \v A} |\v A \>, \quad \phi^{b}_{k} = \sum_{\v A, i} e^{-i 2 \pi \lambda \v k \cdot (\v A + \v b_1)} |\v A + \v b_1 \>
\label{33.2}
\eea
for grids A and B respectively. They satisfy

\bea
\Pi \Pi^{\dagger} \phi^{a,b}_k = \Delta^2 |\sum_i e^{i2 \pi \lambda \v b_i \cdot \v k}|^2 \phi^{a,b}_k \\ \nonumber
\Pi \phi^{a,b}_k = \Delta \sum_i e^{i2 \pi \lambda \v b_i \cdot \v k} \phi^{a,b}_k
\label{34}
\eea
The spectrum and the eigenfunctions are then

\bea
E(\v k)= E_0 \pm \sqrt{\Delta^2 |\sum_i e^{i2 \pi \lambda \v b_i \cdot \v k}|^2 + M^2 } 
\label{35}
\eea
\bea
\psi^{\pm} = C^{\pm} \phi^{a}_k + D^{\pm} \phi^{b}_k, \qquad C^{\pm} = \frac{\pm(E(\v k)-E_0)-M}{\Delta (\sum_i e^{i2 \pi \lambda \v b_i \cdot \v k})} D^{\pm}
\label{36} 
\eea
The degeneracy points of the spectrum for the massless case are $\v k_0 = \pm \frac{1}{2\lambda} (1,-\sqrt{3})$. Expanding around such points one finds

\bea
E(\v k - \v k_0) - E_0 = \pm \sqrt{\Delta^2 k^2 + M^2}
\label{37}
\eea
as expected. One can verify that the Dirac equation is again satisfied.

\subsection{The importance of the tight binding approximation}

We have formulated a theory describing the propagation of waves in hexagonal arrays of resonators with and without deformations. The treatment has been successful in predicting the existence of Dirac points, in compliance with the common knowledge about this system when periodic symmetry is present. However, one may ask whether the tight binding approximation is an essential ingredient for a Dirac-like behaviour of the propagating wave.

To answer this, let us review some of the assumptions we made and the corresponding properties obtained for the two-dimensional Dirac wavefunction. The spin emerges as the probability of the electrons to be located at sites of the triangular sublattice A (spin up) or B (spin down). The momentum (or wave vector) as a conserved quantity came directly from Bloch's theorem, \ie the periodicity of the system. Linearity around degeneracy points (which is essential to produce a Dirac hamiltonian) is a consequence of the hexagonal structure, while the existence of such degeneracy points came from the symmetry under the interchange of the two sublattices. We have seen that the appearence of mass corresponds to the lift of such degeneracy.

In addition to all these properties, one should consider isotropy as a fundamental requirement for the emulation of a free Dirac particle. We claim that rotational symmetry around degeneracy points is a direct consequence of the tight binding approximation, as we shall see. It is well known that rotational symmetry in the Dirac equation demands a transformation of both orbital and spinorial degrees of freedom. It is in the orbital part that we shall concentrate by studying the energy surfaces around degeneracy points beyond the tight binding model.

Let us recall that the transition amplitudes between nearest neighbours (hopping transition) gave rise to the hamiltonian

\bea
H=\sigma_+ \Pi + \sigma_- \Pi^{\dagger}
\label{44}
\eea
where the kinetic operator $\Pi$ could be constructed in terms of translation operators between nearest sites, \ie

\bea
\Pi= \Delta \sum_{i=1,2,3} T_{\v b_i},
\label{45}
\eea
with $T_{\v b_i}$ the one-site translation operator in the direction of $\v b_i$. Since degeneracy points are located using the condition $H \psi_0 = 0$ for a Dirac state $\psi_0$, and given the fact that $[\Pi, \Pi^{\dagger}]=0$, it suffices to impose $\Pi \psi_0 = 0$. The operators $T_{\v b_i}$ are unitary and commute with each other, implying that their eigenvalues can be found simultaneously as $e^{i \lambda \v k \cdot \v b_i}$, where $\v k$ is any real wave vector. Small deviations from degeneracy points (denoted by $\v k_0$) in the form $\v k =\v k_0 + \bfkappa$ give the energy

\bea
E= \Delta | \sum_{i} \exp{(i  \lambda (\v k_0 + \bfkappa) \cdot \v b_i)} | \simeq \Delta \lambda|\bfkappa |,
\label{46}
\eea
which is rotationally invariant in $\bfkappa$. However, one  may try to introduce interactions between sites separated by more than one step under lattice translations. It is clear that a second-neighbour interaction of strength $\Delta'$ modifies the kinetic operator $\Pi$ as

\bea
\Pi= \Delta \sum_{i=1,2,3} T_{\v b_i} + \Delta' \sum_{i=1,2,3} \left( T_{\v a_i} + T_{-\v a_i} \right),
\label{47}
\eea
where the vectors $\v a_i$ have now appeared, connecting a point with its six second neighbors. The energy equation becomes

\bea
E= | \Delta \sum_{i} \exp{(i \lambda \v k \cdot \v b_i)} + \Delta' \sum_{i} 2\cos{( \lambda \v k \cdot \v a_i)} |.
\label{48}
\eea
We expect a deviation of degeneracy points $\v k'_0$, for which $\v k = \v k'_0 + \bfkappa$. Upon linearization of the exponentials in $\bfkappa$ we find the energy

\bea
E \simeq \sqrt{ (\bfkappa \cdot \v u)^2 + (\bfkappa \cdot \v v)^2}
\label{49}
\eea
where the vectors are given by

\bea
\v u = \lambda \Delta \sum_i \cos( \lambda \v k'_0 \cdot \v b_i) \v b_i  \\
\v v = \lambda \Delta \sum_i \sin( \lambda \v k'_0 \cdot \v b_i) \v b_i + 2 \lambda \Delta' \sum_i \sin( \lambda \v k'_0 \cdot \v a_i) \v a_i
\label{50}
\eea 
Thus, the presence of $\Delta'$ yields the energy surfaces (\ref{49}) as cones with elliptic sections whenever $\bfkappa$ is inside the first Brillouin zone. Regardless of how we complete the energy contours to recover periodicity, it is evident that the resulting surfaces are not invariant under rotations around degeneracy points. The circular case is recovered only when $\Delta'=0$, leading to $\v k'_0 = \v k_0$. In this case, the vectors reduce to $\v v = (1,0),\v u = (0,1)$ when $\v k_0$ is the degeneracy point at $(1/2\lambda,0)$.

In summary, extending the interactions to second neighbors has the effect of breaking the isotropy of space {\it around degeneracy points}, which is an essential property of the free Dirac theory.

\subsection{Two-dimensional Dirac oscillator}

Before analyzing lattice deformations, let us recall \cite{nosotros, bermudez} that the two dimensional Dirac oscillator hamiltonian is quite similar to the one dimensional case, except for the replacement $a \mapsto a_R$, where $a_R = a_x + i a_y$ is the chiral (right) anhilation operator in terms of cartesian anhilation operators $a_x, a_y$. The corresponding number operator is a conserved quantity and is given by $N_R = N - L$, \ie the difference between the total number operator and the orbital angular momentum. Since the chiral left operator is absent in the expresion, the spectrum is infinitely degenerate. Eigenfunctions are constructed as a combination of two usual harmonic oscillator functions with defined orbital angular momentum. Our aim is to produce the spectrum for this problem in the hexagonal array, together with a deformation which allows localization of the wave functions around some center.

\begin{figure}
\begin{center}
\includegraphics[scale=0.4]{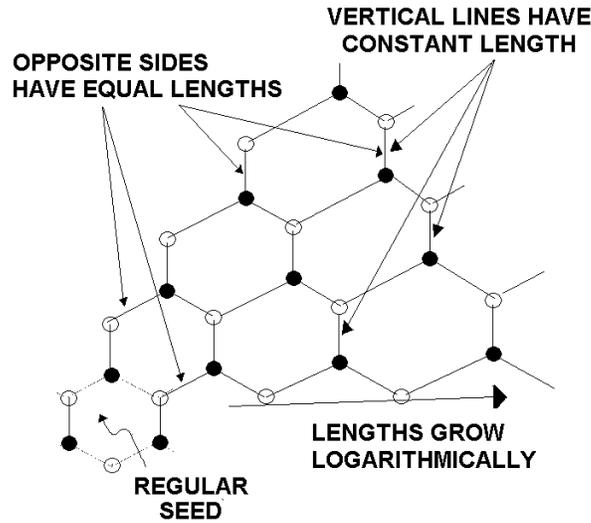}
\end{center}
\caption{\small{Construction of a deformed lattice satisfying the oscillator constraints. Here we show the direction in which the length of hexagons grow as a logarithmic function, departing from a {\it seed\ } represented by a regular hexagon. The cells above are obtained by drawing vectors $\v b_1$ of unit length and then completing the hexagons such that oposite sides are parallel and of the same length.}}
\label{construct1}
\end{figure}

We proceed to deform the lattice through an extension of the kinetic operators, in analogy to the one dimensional case. Let us consider site dependent couplings $\Delta(\v A, \v A + \v b_i)$ connecting the sites labeled by $\v A, \v A + \v b_i$. Again, these are related to distances $d(\v A, \v A + \v b_i)$ between potential wells as $\Delta(\v A, \v A + \v b_i)=\Delta \exp(-d(\v A, \v A + \v b_i)/\Lambda)$. We define the ladder operator

\bea
a_R= \sum_{\v A, i} \Delta(\v A, \v A + \v b_1) \left( |\v A \> \< \v A + \v b_i - \v b_1 | + |\v A + \v b_1 \> \< \v A + \v b_i | \right)
\label{38}
\eea
and impose $[a_R,a_R^{\dagger}]=\omega \Delta$. After some algebra, one can prove that this leads to the conditions

\bea
 \Delta(\v A, \v A + \v b_1) = \Delta,
\label{con1}
\eea

\bea
\Delta^2(\v A, \v A + \v b_2) + \Delta^2(\v A + \v b_2, \v A + \v b_2 - \v b_3) = \label{con2.1} \\ 
\Delta^2(\v A + \v b_1, \v A + \v b_1 - \v b_3) + \Delta^2(\v A + \v b_1 - \v b_3, \v A + \v b_1 + \v b_2 - \v b_3),
\nonumber
\eea

\bea
 \Delta^2(\v A, \v A + \v b_2) + \Delta^2(\v A, \v A + \v b_3) = \label{con3.1} \\ 
\Delta^2(\v A + \v b_1, \v A + \v b_1 - \v b_3) + \Delta^2(\v A + \v b_1, \v A + \v b_1 - \v b_2) + \omega \Delta.
\nonumber
\eea
The vector $\v b_1$ in the first equation was chosen arbitrarily, but due to symmetry a choice of $\v b_2$ or $\v b_3$ would be equivalent. Due to the coordination number three, we obtain three relations rather than the two of the one-dimensional case. As complicated as the recursion relations may seem, one can easily construct a lattice reproducing them consistently. The relations (\ref{con1}) and (\ref{con2.1}) establish an equality between the lengths of opposite sides of a given hexagon. The relation (\ref{con3.1}) containing $\omega$ gives the deformation and can be split in two parts

\bea
\Delta^2(\v A, \v A + \v b_2) - \Delta^2(\v A + \v b_1, \v A + \v b_1 - \v b_2) =  \omega \Delta \sin^2 \theta
\label{39.9}
\eea

\bea
\Delta^2(\v A, \v A + \v b_3) - \Delta^2(\v A + \v b_1, \v A + \v b_1 - \v b_3) = \omega \Delta \cos^2 \theta
\label{40}
\eea
where $\theta$ is an arbitrary angle. Having chosen previously the privileged direction $\v b_1$, the angle $\theta$ determines the relative stretching between directions $\v b_2$ and $\v b_3$ on the seminal cell. The choice $\theta=0$ produces deformations only in one direction of the lattice ($\v b_3$), resembling the one dimensional case discussed above. This leads to a logarithmic law for the deviation distance similar to (\ref{12}). Logarithmic stretching will hold for an arbitrary angle $\theta$. To construct the grid in the general case, one starts with a regular hexagon as a seed and completes the scheme in figure \ref{construct1} by extending lines of equal length in the direction of $\v b_1$. Then, one completes the hexagons by drawing parallel lines for the opposite sides as shown also in the figure \ref{construct1}. Hexagonal cells satisfy the recursion relations above trivially.

A restriction to a finite dimensional space occurs in a way similar to the one dimensional case. The maximum number of levels is $N_{max} = [|\frac{\Delta}{\omega}|]$. 

Considering $a_R$ as the chiral operator restricted to this finite dimensional space, the resulting hamiltonian of this problem is

\bea
H= E_0 + \sigma_3 M + \sigma_+ a_R + \sigma_- a_R^{\dagger}.
\label{42}
\eea
As the hamiltonian (\ref{42}) is formally identical to that in the one-dimensional case, we find the eigenvalues

\bea
E^{\pm}(N_R+1)= E_0 \pm \sqrt{ \omega \Delta (N_R+1) + M^2 }, \qquad 0 \leq N_R \leq \Delta/\omega, \\ \nonumber
E(0)= E_0-M.
\label{43}
\eea
The shape of the eigenfunctions is obtained by solving $a_R \phi_{0} = 0$ and applying the raising operators similar to appendix A. 

The hamiltonian does not depend on the left operators $a_L,a^{\dagger}_L$ where $a_L = (a_R)^*$. In the full space this would imply a Landau electron-like infinite degeneracy. The degeneracy does not occur for a fixed array of resonators, but it can be interpreted to reflect the arbitray choice of $\theta$ if we consider the hypothetical use of an ensemble of arrays for all angles $\theta$. 

Note that there is a physical limitation to the allowed degree of distortion which results from the fact that the nearest neighbours can change. Before this happens, the coupling of potential wells can no longer be dominated by the original three nearest neighbours.

\begin{figure}[h!]
\begin{center}
\includegraphics[scale=0.9]{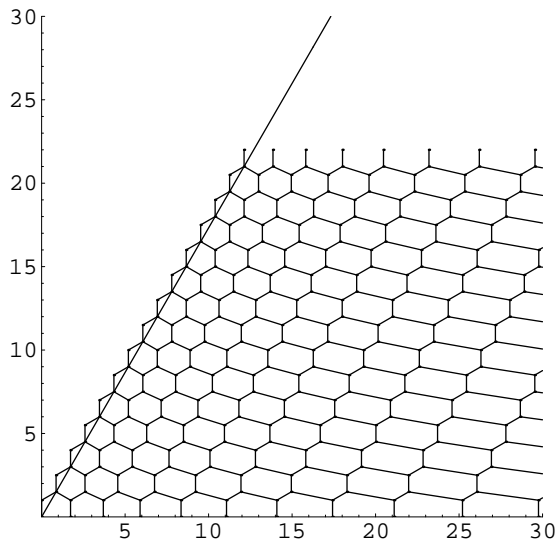}
\end{center} 
\caption{\small{Case 1: Two dimensional lattice obtained by setting $\theta=0$, $\lambda=1$, $\omega=1/15$. We start by deforming the vector $\v b3$ near the seminal cell at the origin. For the rest of the lattice, we follow the recurrence relations and the construction indicated in the text. Periodicity appears in the direction $\v b1-\v b2$ indicated with a line at 60 degrees and passing through the origin.}}
\label{a}
\end{figure}

\begin{figure}[h!]
\begin{center}
\includegraphics[scale=0.9]{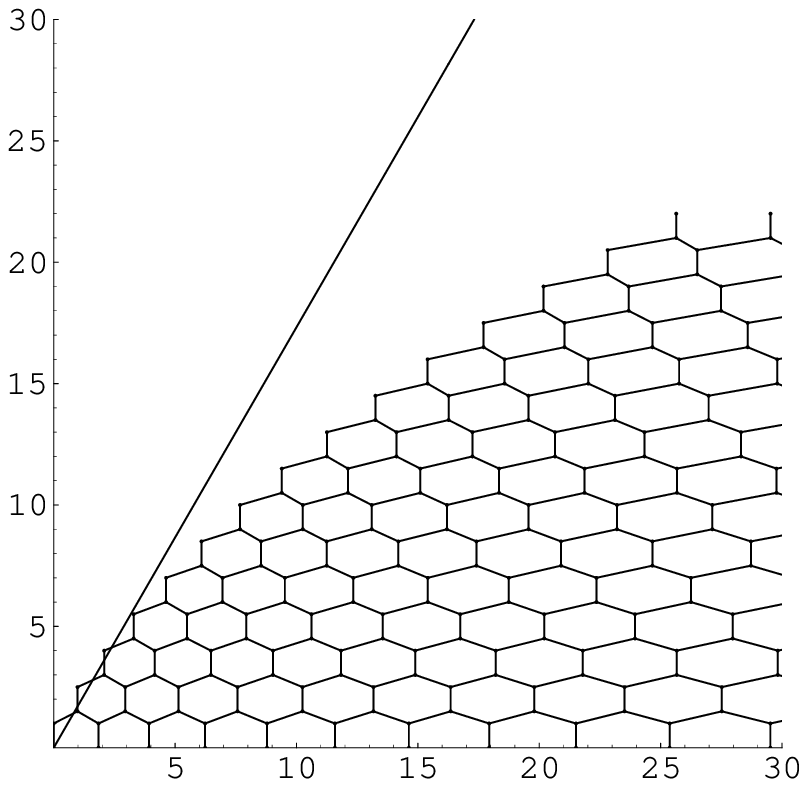}
\end{center} 
\caption{\small{Case 2: Two dimensional lattice obtained by increasing the angle $\theta=\pi/4$. As before, $\lambda=1$, $\omega=1/15$. In this case both vectors $\v b2$ and $\v b3$ are deformed near the seminal cell at the origin. The rest of the lattice is constructed by using the recurrence relations and completeting the hexagons such that opposite sides have equal length. There is no periodic symmetry in this case.}}
\label{b}
\end{figure}

\begin{figure}[h!]
\begin{center}
\includegraphics[scale=0.9]{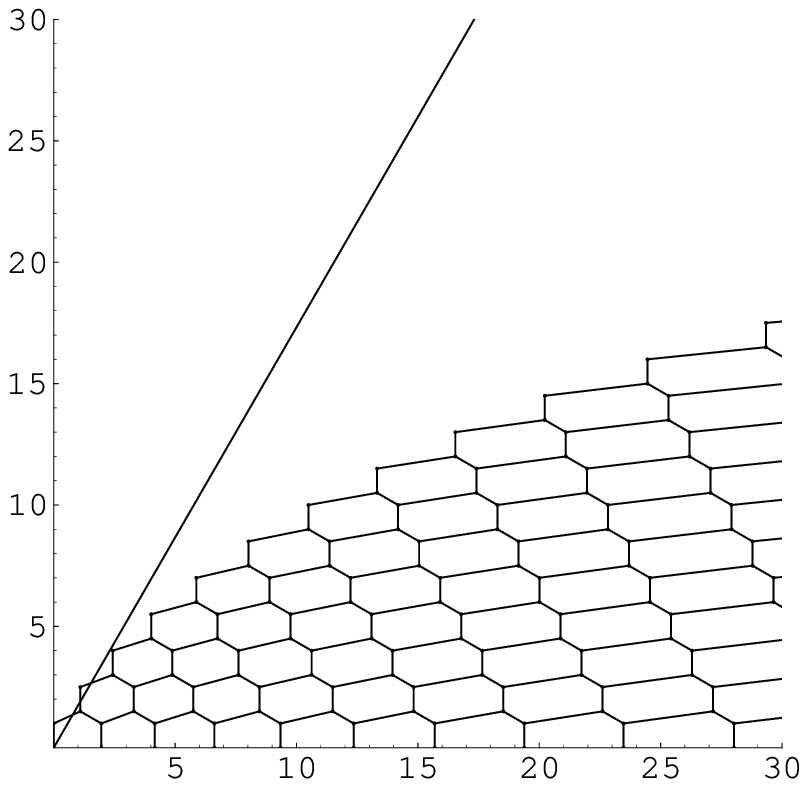}
\end{center} 
\caption{\small{Case 3: Two dimensional lattice obtained by setting $\theta=\pi/2$. As before, $\lambda=1$, $\omega=1/15$. In this case, the vector $\v b2$ is deformed near the regular cell at the origin. The rest of the lattice is constructed by using the recurrence relations and completeting the hexagons such that opposite sides have equal length. There is no periodic symmetry in this case.}}
\label{c}
\end{figure}

In figures \ref{a}, \ref{b}, \ref{c} we give some realizations of lattice deformations following the procedure indicated previously. The examples are related to the choices of $\theta=0,\pi/4,\pi/2$, which determine the vectors to be deformed with a logarithmic law near the x axis of the graphs. 

\section{Experimental implementation in microwave cavities}

To emulate Dirac-like equations on two dimensional arrays of resonators, we have to generate a scalar field. For electromagnetic fields, this can be achieved in a 2D metallic cavity which supports two types of independent modes: transverse magnetic (TM) mode, $\psi(\v r)=E_z(\v r)$, and transverse electric (TE) mode, $\psi(\v r)=H_z(\v r)$, $\v r$ lying in the plane of the cavity. For a cavity of height $h$ (in the $z$-direction) and for frequencies $\nu<c/2h$, the scalar field obeys the Helmholtz equation
\begin{equation}
-\nabla^2 \psi(\v r) = k^2\psi(\v r).
\end{equation}
For the top and bottom plates, it fulfills Dirichlet or Neumann boundary conditions, for TM and TE modes respectively. For cavities with a typical horizontal extension of a tens of centimeters and a  typical height of a few millimeters, the frequency range of interest lies in the domain of microwaves. Such cavities constitute paradigmatic examples for a variety of phenomena in open and closed two-dimensional systems as discussed in the introduction.
 
The experimental set-ups using chaotic microwave cavities adopt various configurations depending on the specific studies they are intended for, but the technique to feed microwaves into the cavity and to collect the signal of interest is ubiquitous. The central device is the network analyzer which performs emission and lock-in detection of microwaves from a few tens of MHz to a few tens of GHz. The cavity is linked to the network analyzer through flexible coaxial 50\,$\Omega$-cables connected to monopolar antennas whose central conductor penetrates into the cavity. Usually, several antennas are dispatched over the top and/or bottom plates.  For a measurement, only one antenna at a time is used as a microwave emitter and another (in transmission) or the same (in reflection) as a receiver. The other unused antennas are terminated by 50\,$\Omega$ loads so that all antennas behave the same way regarding the losses they imply. The measurements are given in terms of scattering coefficients which form the complex $S$-matrix

\bea
 S=\left(\begin{array}{cc} S_{11} & S_{12}\\S_{21} & S_{22}\end{array}\right), 
\label{smatrix}
\eea
where $S_{11}$ (resp.  $S_{22}$) measures the reflection on port 1 (resp. 2) and $S_{12}$ (resp.  $S_{21}$) measures the transmission from port 2 (resp. 1) to port 1 (resp. 2). All the measurements are performed after a proper calibration to get rid of any parasitic influence of cables and connectors and even of the analyzer itself. 

In a microwave cavity, a varying potential can be obtained by introducing substances of a varying permitivity $\epsilon(\v r)$. The wave equation then reads:
\begin{equation}
\label{waveeq}
\big[-\nabla^2 +\big(1-\epsilon(\v r)\big)k^2\big]\psi(\v r) = k^2\psi(\v r).
\end{equation}
Note that the effective potential $\tilde{V}(\v r)=\big(1-\epsilon(\v r)\big)k^2$ is energy-dependent. This does not preclude the quantum-classical analogy. 

Recently, one of us developed experiments implementing equation (\ref{waveeq}) in a disordered microwave cavity\cite{Lau07}. The physical phenomenon put under scrutiny was Anderson (or strong) localization (see \cite{Lag09} for a recent survey of this prolific domain). A network analyzer Rohde \& Schwarz ZVA-24 was used in a frequency range from 1\,GHz to 10\,GHz. The disordered potential was introduced through 200 dielectric cylinders (Temex-Ceramics, E2000 series) of high dielectric permitivity ($\epsilon=37$) and low loss (quality factor $Q=7\,000$ at 7\,GHz). Their height fitted that of the cavity, 5\,mm, and several diameters ranging from 6\,mm to 8\,mm were used. The disorder was numerically generated and the scatterers were precisely positioned. In this experiment the central conductor of the antennas was perpendicular to the plane of the cavity. Thus, the TM polarization of the electromagnetic field was selected. \cite{Lau07}\cite{Seb06}.

The same dielectric cylinders, used for the localization experiments, can be arranged in periodic or other ordered patterns. In the domain of optics, periodic structures in semiconductor materials are widely used to obtain particular transport properties of guided light. Thanks to photonic crystals (also called photonic band gap materials) the technology of photons conspire to supplant the one of electrons in domains such as communication and information technologies, computing and sensing \cite{Lou05}. Microwave cavity experiments, which are definitely intended for more fundamental issues, bear the benefits of their versatility. They allow, for example, distortions destroying the periodicity of the potential such as the ones described in sections 2 and 3 for emulating Dirac oscillators. 
 
As emphasized in section 3.2, we have to apply the tight binding condition. In the localization experiment, the field filled all the cavity, TM polarization being supported inside and outside the resonators. For the experiments we proposed with ordered structures, the requirements are quite different. In order to transmit the energy efficiently into the cylinders, we use in-plane antennas, and consequently TE polarization. Due to the wavelength reduction by a factor $\sqrt{\epsilon}\simeq 6$ inside the dielectric material, TE modes can be excited into the scatterers above 5\,GHz, while 30\,GHz is required in air between the resonators. In the proposed frequency range, resonators support modes which decay evanescently in the surrounding space. This implies exponential decay outside the resonators. For a dielectric cylinder of 8\,mm in diameter, the first TE mode appears at 6.66\,GHz. We experimentally checked that this mode is isotropic: $\psi(\v r)\sim J_0(\v r)$, thus enabling isotropic coupling between resonators. We studied the range of the coupling and its spatial dependence. Measurements were done with 2 and 3 resonators equally spaced on a line, and 6 resonators placed at the vertices of a hexagon (a benzene-like structure). For all configurations, we observed an exponentially decreasing coupling with a characteristic length of $400\,$m$^{-1}$. The 3-cylinder and benzene measurements clearly established that the second-neighbor coupling is negligible for distances between the centers of the scatterers above 10\,mm. This gives a large range of the coupling constant for which the tight-binding model is fulfilled. All these experimental results will be published in a forthcoming paper focused on transport properties in graphene-like structures\cite{Bar09}.

We have thus established that we can meet the conditions for the emulation of the Dirac oscillator and related problems can be met with arrays of dielectric microwave oscillators between two condcuting plates. These conditions will also allow to emulate other Dirac operators corresponding to gyroscopes, disordered systems, etc. Note though that it will be difficult to emulate a Dirac hydrogen atom as the distances between resonators should become extremely small, causing conflict with the diameter of the discs. 

\section{Conclusions and outlook}

We have proposed that a wide class of relativistic equations of the Dirac type can be implemented by arrays of potential wells. Such quantum systems can be well approximated by tight binding hamiltonians if sufficiently fast exponential decay of the wave functions in the classically prohibited region is ensured. This idea was especifically implemented for the Dirac oscillator. Furthermore we have shown that we can implement such a situation experimentally with arrays of dielectric resonators. The use of TE modes and the conducting plates below and above of the resonators ensures at appropriate frequencies well isolated resonances in the resonators and exponential decay outside. We have thus revealed a practical way to emulate Dirac-like equations for both massive and massless particles using classical wave systems.

While we presented an emulation for the Dirac oscillator in one and two dimensions, it is clear that our algebraic treatment allows other possibilities. The most promising example is the emulation of Dirac gyroscopes \cite{yo} as the number of states in this case is finite to begin with due to conservation of total angular momentum. This forces its realization on finite grids without approximations in that respect. The relativistic hydrogen atom, on the other hand presents obvious difficulties regarding the steep potential needed near its singularity and the nearly flat potential at large distances requires separations too small for the radii of the resonators used. Some intermediate region of Rydberg states could possibly be emulated. Leaving the realm of integrable systems, it might be interesting to introduce random small perturbations in the positions of the wells. This might mimic some properties of random matrix Dirac operators \cite{verbaarschot}.

\appendix

\section{Eigenfunctions}

\setcounter{section}{1}

\begin{figure}[h!]
\begin{center}
\includegraphics[scale=0.4]{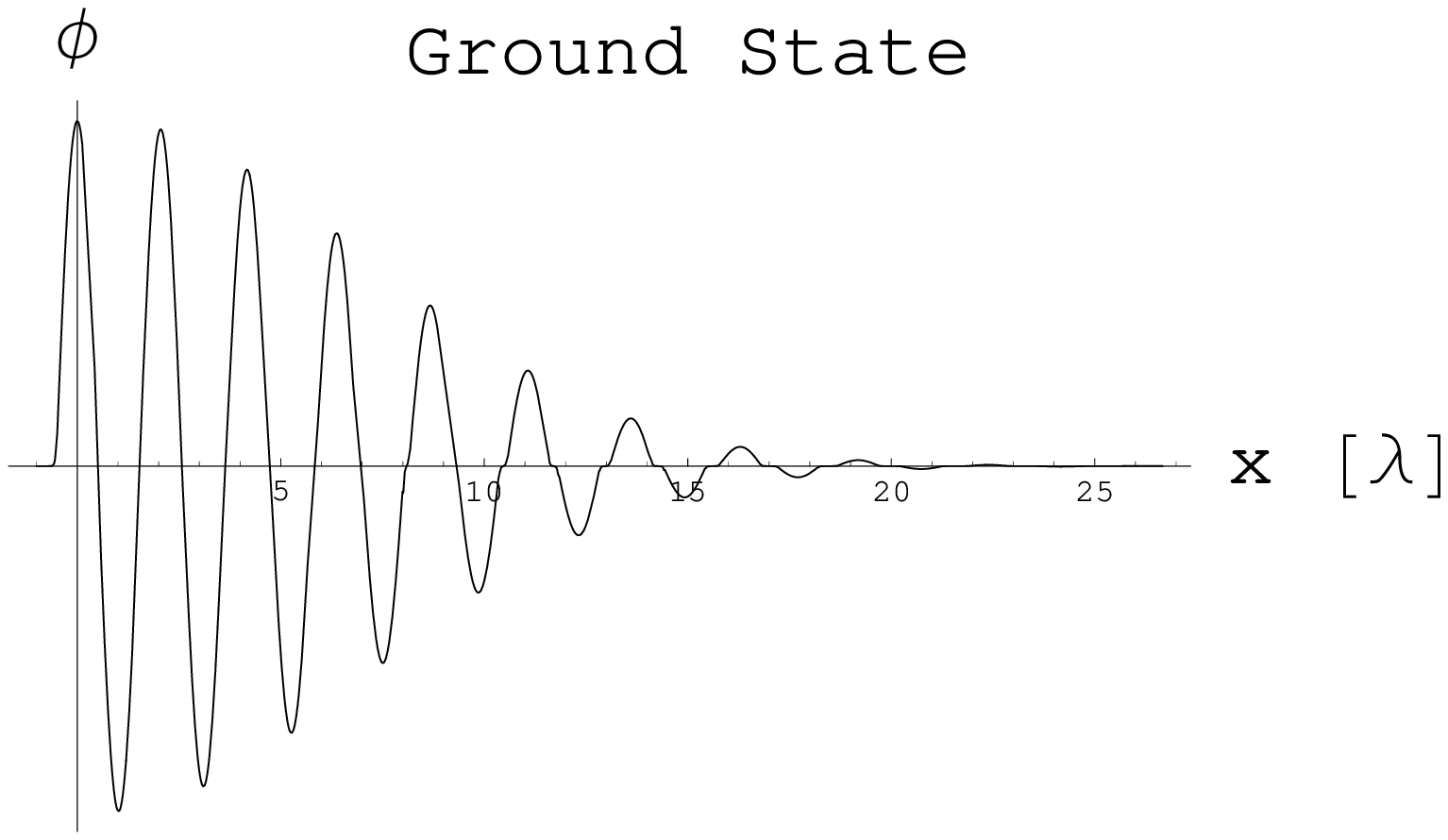}
\end{center}
\caption{\small{Ground state wavefunction in position space (in units of $\lambda$). The localized wave functions are chosen to be the groundstate of individual wells with energy $\alpha=\beta=1$. The width of the wells is $2\lambda/3$. The parameters of the lattice are $n_{max}=20$, $\omega=1/20$ and $\Delta=1$. A gaussian envelope is visible. The signs alternate from site to site.}}
\label{gs1}
\end{figure}

\begin{figure}[h!]
\begin{center}
\includegraphics[scale=0.4]{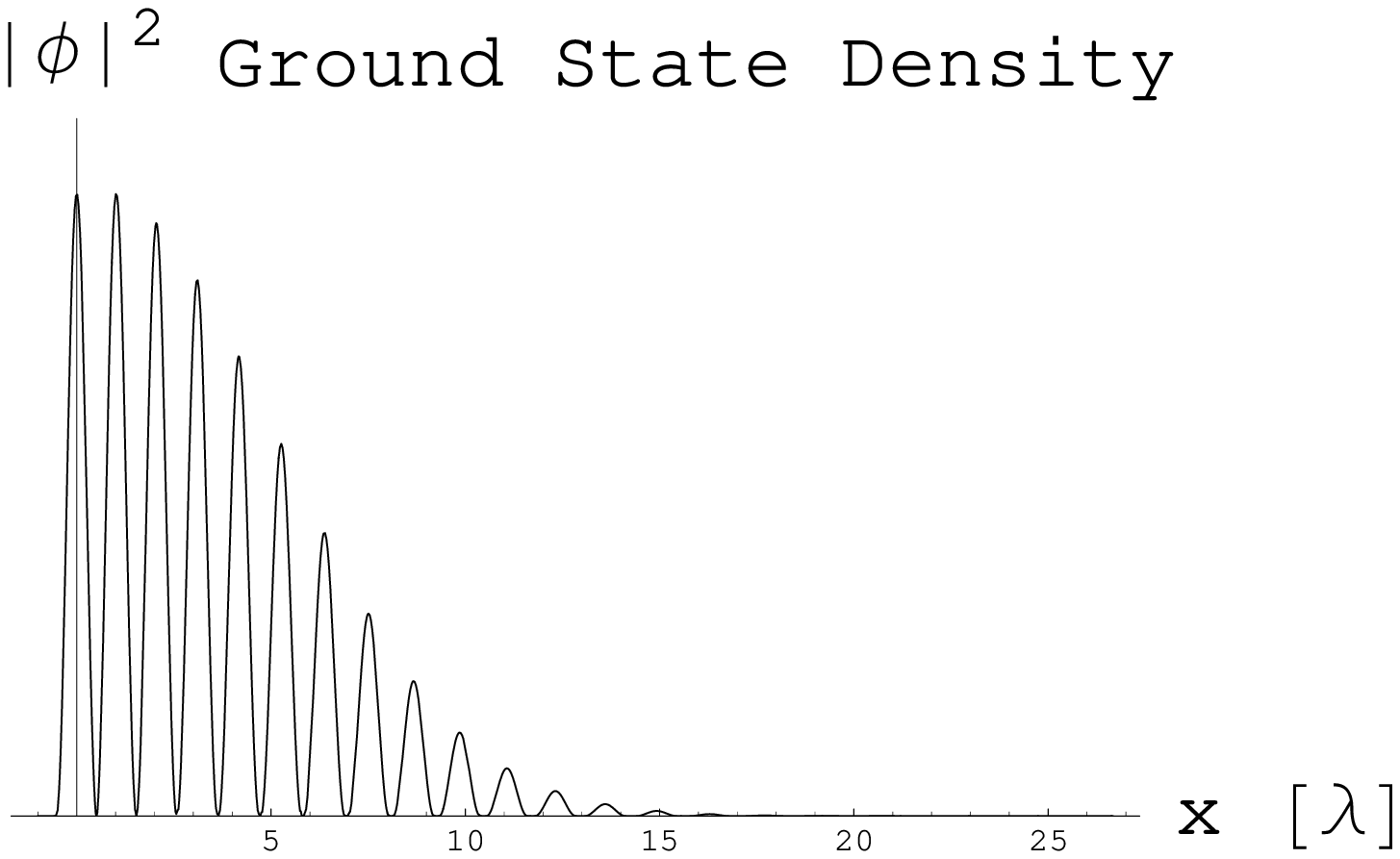}
\end{center}
\caption{\small{Ground state density as a function of position in units of $\lambda$. The localized wave functions are the groundstates of individual wells with energy $\alpha=\beta=1$ and width $2\lambda/3$. The parameters of the lattice are $n_{max}=20$, $\omega=1/20$ and $\Delta=1$. A gaussian envelope is visible.}}
\label{gs2}
\end{figure}

We first determine the ground state $\phi_0$. We write

\bea
\phi_0 = \sum_{m=0}^{n_{max}} f_0(m) |m\>
\label{25}
\eea
and use $a \phi_0 = 0$ to obtain the recurrence equation

\bea
(\sqrt{\Delta^2- (n_{max}-m) \omega \Delta}) f_0(m+1) + \Delta f_0(m) = 0
\label{26}
\eea
with solution

\bea
f_0(m) = (-1)^m C_0 \prod_{j=0}^{n_{max}-m} \frac{\Delta}{\sqrt{\Delta^2-(n_{max}-j) \omega \Delta}} \\ \nonumber
C_0 = \left( \sum_{m=0}^{n_{max}} \Pi_{j=0}^{n_{max}-m} \frac{\Delta^2}{\Delta^2-(n_{max}-j) \omega \Delta} \right)^{-1/2}.
\label{27}
\eea
Applying raising operators we obtain

\bea
\phi_n = \frac{(a^{\dagger})^n}{\sqrt{(\omega \Delta)^n n!}} \phi_0
\label{27.1}
\eea
The space-dependent wave functions $\phi_n(\v r)$ are obtained by defining a specific form of the localized functions $\xi_A, \xi_B$. For definitness, we choose these functions to be the ground state of a deep square well. The ground state $\phi_0(\v r)$ and the density
$|\phi_0(\v r)|^2$ are shown in figures 
\ref{gs1} and \ref{gs2}.

\ack
F. M. acknowledges the kind hospitality of CICC, UNAM. We thank U. Kuhl, C. Lewenkopf, S. Barkhofen and T. Tudorowskiy for useful discussions. This work was financially supported by UNAM under project PAPIIT IN-107308 and by CONACYT under projects 79613 and 57334.

\end{document}